\documentclass[preprint,showpacs,preprintnumbers,amsmath,amssymb]{revtex4}
\usepackage[utf8]{inputenc}

\usepackage{graphicx}
\usepackage{psfrag}
\usepackage{amsmath}

\begin{document}

\preprint{TUD-ITP-TQO/04-2010-V100303}

\title{Time-dependent Semiclassics for Ultracold Bosons}

\author{Lena Simon}
     \affiliation{Institut f\"{u}r Theoretische Physik,
Technische Universit\"at Dresden, D-01062 Dresden, Germany}
\author{Walter T. Strunz}
   \affiliation{Institut f\"{u}r Theoretische Physik,
Technische Universit\"at Dresden, D-01062 Dresden, Germany}

\date{\today}

\begin{abstract}
We study the out-of-equilibrium dynamics of ultracold bosons in a double- and triple-well potential within the 
Bose-Hubbard model by means of the semiclassical Herman-Kluk propagator and compare the results to the 
frequently applied ``classical dynamics'' calculation in terms of the truncated Wigner approximation (TWA). For the double-well system we find the 
semiclassical results in excellent agreement with the numerically exact ones, while the TWA is not able to 
reproduce any revivals of the wave function. 
The triple-well system turns out to be more difficult to handle due to 
the irregularity of the corresponding classical phase space. Here, deviations of the TWA from the exact dynamics 
appear even for short times, while better agreement is obtained using the semiclassical approach presented in this article.

\end{abstract}

\maketitle

\section{Introduction}
The dynamics of interacting many-body quantum systems is an active 
field of research. 
There are many open questions concerning the occurrence of relaxation and the corresponding observables, how the initial state and system parameters influence the dynamics and on which timescales 
a possible thermalization might appear. 
While in equilibrium statistical mechanics established approaches exist for 
describing the state of a system, finding widely applicable computational methods to gain deeper insight into interacting many-body quantum dynamics is an important task \cite{Polk11}. 
\\
\\
Ultracold bosonic gases in optical lattices provide a perfect playground to investigate out-of-equilibrium 
many-body quantum dynamics \cite{Kenn12}, due to being easily controlled and varied with time in experiments \cite{Trotz12}. 
From the theoretical point of view, those kinds of systems can be described in the framework 
of the famous Bose-Hubbard model, as proposed in Ref. \cite{Jak98}. However, 
solving the respective exact quantum dynamics is possible for only a few particles, even 
for only weakly interacting gases, thus making approximations necessary. 
\\
\\
One of the main theoretical tools in the field of ultracold bosons is the Gross-Pitaevskii equation, 
a field equation for the Bose-Einstein condensate wave function in terms of a mean-field approximation, 
providing accurate results on limited timescales for low temperatures and for large numbers of particles $N$. 
However, replacing the field operators by a \textit{c}-number field means that some truly quantum phenomena 
(such as wave packet revivals) cannot be described and the mean-field approach quickly ceases to be valid - especially for strong interactions \cite{String03}. 
Hence, this article deals with the question whether a \textit{semiclassical} treatment can improve the classical mean field 
Gross-Pitaevskii approach and existing extensions thereof. 
\\
\\
Semiclassical methods are widely applied in the energy picture, in terms of the WKB and related approximations \cite{Berry72}. 
On the other hand, time-dependent semiclassical propagators have a long history, going back to Van Vleck \cite{VV28}, especially in the 
field of molecular dynamics. Besides the Van-Vleck-Gutzwiller propagator \cite{Gutz90}, which is based on a classical 
boundary value problem, the so-called Herman-Kluk (HK) propagator is frequently applied \cite{Frank}, providing the advantage of relying on 
a classical initial value problem, thus avoiding the expensive root-finding task. 
\\
The propagator turns out to depend on classical information only, similar to the popular Truncated Wigner Approximation (TWA), which is employed 
to describe the full quantum field dynamics as well. However, the semiclassical propagator goes beyond the TWA by considering the coherences of the contributions of the classical trajectories, without implicating major additional expenses. 
In the following we will refer to the TWA as the ``classical approximation''. 
\\
\\
To test the applicability of the HK propagator in the field of ultracold bosons, we consider as model systems ultracold bosons in a double- and triple-well potential, which 
allows for an exact calculation of the dynamics in the Bose-Hubbard framework for a comparison of the results. 
For the former, experiments exist which confirm the two-mode picture \cite{Alb05,Zib10} on the one hand, and a number of articles dealing with the discussion 
of the consequences of the mean-field approximation and many-body corrections \cite{Var01,Ang01} on the other. 
Furthermore, (time-independent) semiclassical methods have been applied successfully to the two-mode model, 
with WKB methods leading the way \cite{Fran00,Gra07,Sh08,Chu10,Nis10,Itin11,Paw11,Sim12}, showing that semiclassical 
methods are a promising tool. In this article, we want to discuss the application of the (explicitly time-dependent) 
semiclassical Herman-Kluk propagator for ultracold bosons. 
Semiclassical approximations are expected to be valid in the limit $\hbar \rightarrow 0$, or more precisely when $\hbar$ is small 
compared to characteristic actions of the system. For the quantum many-body systems of interest here, one usually starts from 
the second-quantized many-body Hamiltonian for ultracold bosons 
\begin{equation}
\label{eq_many_body_ham}
 \hat H = \int \left(-\frac{\hbar^2}{2m} |\nabla \hat \psi(\boldsymbol{r})|^2+V(\boldsymbol{r})\hat \psi ^\dagger(\boldsymbol{r}) \hat \psi(\boldsymbol{r}) 
+ \frac{g}{2} \hat \psi^\dagger(\boldsymbol{r})  \hat \psi^\dagger(\boldsymbol{r})  \hat \psi(\boldsymbol{r})  \hat \psi(\boldsymbol{r})\right) \mathrm{d} \boldsymbol{r}
\end{equation}
with the external potential $V(\boldsymbol{r})$ and the interaction parameter $g=\frac{4\pi\hbar^2a}{m}$ (depending on the $s$-wave scattering 
length $a$). 
By normalizing the bosonic field operators to the square root of the particle number in the Hamiltonian (\ref{eq_many_body_ham}) 
it becomes clear that semiclassical methods are expected to be valid in the limit
\begin{equation}
 \label{class_limit}
N \rightarrow \infty \ , \ \ g \rightarrow 0 \ \  \text{with} \ \ gN= \text{const} \ , 
\end{equation}
the so-called classical limit \cite{Lieb00}, in which the Gross-Pitaevskii ground state is exact. 
Indeed, this limit is reasonable in the field of ultracold bosonic gases, 
since most of the experiments involve many weakly interacting particles. 
\\
\\
In this article, we first review the semiclassical Herman-Kluk propagator in detail as well as the commonly applied 
truncated Wigner approximation for a comparison between the semiclassical and the classical methods. 
Afterwards, we discuss the double-well system and its semiclassical dynamics, finding good agreement with the exact dynamics for almost all 
parameter regimes. We show that the semiclassical results outmatch the classical mean field and TWA results clearly. 
Finally, the last section deals with the dynamics of ultracold bosons in a triple-well potential. 
Again, the semiclassical results are compared to the exact dynamics and the TWA. We come to the conclusion that 
semiclassics is superior to the TWA, but both approaches are in accordance with the exact dynamics on only fairly short timescales, 
because of the mixed phase space of the triple-well system.

\section{The semiclassical Herman-Kluk propagator}

The time evolution of an initial wave function $\psi(\boldsymbol{r}_i,0)$ can be expressed 
by means of the quantum mechanical propagator $K$ according to
\begin{equation}
 \psi(\boldsymbol{r}_f,t)=\int \mathrm{d}\boldsymbol{r}_i K(\boldsymbol{r}_f,t;\boldsymbol{r}_i,0) \psi(\boldsymbol{r}_i,0) \ . 
\end{equation}
If the typical actions involved in the system are large compared to $\hbar$, the 
propagator can be approximated semiclassically employing fixed width (so-called frozen) Gaussian wave packets $\langle \boldsymbol{r}|\boldsymbol{z}(\boldsymbol{r},\boldsymbol{p})\rangle$, 
centered around the phase space points $\boldsymbol{r}$ and $\boldsymbol{p}$. 
The resulting propagator was developed by Herman and Kluk \cite{HK84}, based on previous work done by Heller \cite{Hel81}, 
and brought back to the center of attention by Kay \cite{Kay94,Kay942}. The so-called 
Herman-Kluk propagator applied to an initial Gaussian state $|\boldsymbol{z}_0\rangle$ reads \cite{Frank}
\begin{equation}
\label{eq_hk}
 K(\boldsymbol{r}_f,t;\boldsymbol{z}_0,0)=\int \frac{\mathrm{d}^Dp\mathrm{d}^Dq}{(2\pi\hbar)^N}\langle \boldsymbol{r}_f | \boldsymbol{z}(t)\rangle R(\boldsymbol{q},\boldsymbol{p},t)\exp\left(\frac{i}{\hbar}S(\boldsymbol{q},\boldsymbol{p},t)\right) 
\langle \boldsymbol{z} | \boldsymbol{z}_0 \rangle \ , 
\end{equation}
where $D$ denotes the degrees of freedom of the system. 
It is integrated over phase space points $(\boldsymbol{q},\boldsymbol{p})$ being the initial conditions of the propagated classical trajectories $(\boldsymbol{q}(t),\boldsymbol{p}(t))$. 
The normalized Gaussian states in position space are
\begin{equation}
 \langle \boldsymbol{r}|\boldsymbol{z}\rangle = \left(\frac{\gamma}{\pi}\right)^{N/4} \exp \left(-\frac{\gamma}{2}(\boldsymbol{r}-\boldsymbol{q})^2+\frac{i}{\hbar}\boldsymbol{p}(\boldsymbol{r}-\boldsymbol{q})\right) \ ,
\end{equation}
where the parameter $\gamma$ determines the width of the Gaussians, and the overlap of two states can be calculated analytically to
\begin{equation}
\label{eq_overlap}
 \langle \boldsymbol{z} |\boldsymbol{z}_0 \rangle = \exp\left(-\frac{\gamma}{4}(\boldsymbol{q}-\boldsymbol{q}_0)^2+\frac{i}{2\hbar}(\boldsymbol{q}-\boldsymbol{q}_0)(\boldsymbol{p}+\boldsymbol{p}_0)-\frac{1}{4\gamma\hbar^2}
(\boldsymbol{p}-\boldsymbol{p}_0)^2\right) \ , 
\end{equation}

if the widths of the two Gaussians $\gamma$ are equal. The phase of the integrand in Eq. (\ref{eq_hk}) depends on the classical action $S=S(\boldsymbol{q},\boldsymbol{p},t)=\int L \mathrm{d} t$, with the classical Lagrangian 
$L=T-V$, which is integrated along the propagated trajectories. 
Finally the complex HK prefactor reads
\begin{equation}
\label{eq_HK_pref}
 R(\boldsymbol{q},\boldsymbol{p},t)=\det\left[\frac{1}{2}\left(\boldsymbol{m}_{11}+\boldsymbol{m}_{22}-i\hbar\gamma \boldsymbol{m}_{21}-\frac{1}{i\hbar\gamma}\boldsymbol{m}_{12}\right)\right]^{1/2} \ ,
\end{equation}
including elements of the so-called monodromy matrix 
\begin{equation}
\label{eq_monodromy}
 \mathbf{M}=\begin{pmatrix}\boldsymbol{m}_{11} & \boldsymbol{m}_{12} \\ \boldsymbol{m}_{21} & \boldsymbol{m}_{22} \end{pmatrix} = \begin{pmatrix} \frac{\partial \boldsymbol{p}(t)}{\partial \boldsymbol{p}} 
                                                                                                                                                 & \frac{\partial \boldsymbol{p}(t)}{\partial \boldsymbol{q}} \\
                                                                                                                                                 \frac{\partial \boldsymbol{q}(t)}{\partial \boldsymbol{p}}
                                                                                                                                                  & \frac{\partial \boldsymbol{q}(t)}{\partial \boldsymbol{q}}
                                                                                                                                  \end{pmatrix} \ ,
\end{equation}
being solutions of linearized Hamilton equations of motion which can be calculated along with the trajectories. 
The HK prefactor weights the contribution of particular trajectories depending on their stability with respect to slightly different initial conditions. 
Unlike in Ref. \cite{Juan13}, where a semiclassical propagator based on a boundary value problem is applied to the dynamics of ultracold bosons in a lattice, 
the advantage of the HK propagator is that the classical information appears in terms of an initial value problem.  

\subsection{Truncated Wigner Approximation}
The TWA constitutes another widely used method to simulate the dynamics of out of equilibrium BECs. 
Its application to BECs was suggested in Ref. \cite{Steel98} as an alternative to go beyond Gross-Pitaevskii theory. 
The method is based on the Wigner representation of the density operator whose time evolution is given by the Von-Neumann equation. For an ultracold bosonic gas the time evolution 
of the Wigner function contains first- and third-order derivatives with respect to the classical field \cite{Steel98,Sin02}, such that a direct integration of the 
equation of motion is impractical and difficult. To avoid this difficulty, the higher derivatives are typically truncated, leading to a classical Liouville equation for the 
time-evolution of the Wigner function \cite{Steel98,Sin02}. So, each classical field of an ensemble of fields, which samples the Wigner function of the initial density operator of the gas, 
is evolved by means of the Gross-Pitaevskii equation. Even though the classical equation of motion is completely deterministic, quantum noise is still included in the initial state represented by 
a distribution of classical fields. 
The TWA turned out to be a good approximation for large numbers of particles and to be exact in the classical limit (\ref{class_limit}) \cite{Blak09}, 
provided the corresponding Wigner function is sufficiently smooth. 
If not, the third order derivatives can become relevant. Then semiclassical methods presented here in form of the HK propagator, and also relying on ``classical information'' only become preferable. 
To which extent the semiclassical methods outmatch the TWA is investigated on the basis of ultracold bosons in a double-well and a triple-well potential in the following sections.

\section{Ultracold Bosons in a double-well potential}
A Bose-Einstein condensate in a symmetric double-well potential can be described for low temperatures by means of the two-mode approximation \cite{Gat07}. 
Starting from the many-body Hamiltonian (\ref{eq_many_body_ham}), after some approximations the corresponding second-quantized many-body two-site Bose-Hubbard Hamiltonian reads
\begin{equation}
\label{eq_BH}
 \hat H_{BH}=-T(\hat a_1^\dagger \hat a_2 + \hat a_2^\dagger \hat a_1)+U(\hat a_1^\dagger \hat a_1^\dagger \hat a_1 \hat a_1 + \hat a_2^\dagger \hat a_2^\dagger \hat a_2 \hat a_2) \ ,
\end{equation}
with creation and annihilation operators $\hat a_i^\dagger$ and $\hat a_i$ for a boson in the $i$th potential well, which obey the usual bosonic commutation rules $[\hat a_i,\hat a_j^\dagger]=\delta_{ij}$. 
$T$ denotes the tunnelling amplitude, which depends on the barrier height \cite{Jak05}, and $U$ measures the on-site two-body interaction strength depending on the $s$-wave scattering length. 
The Bose-Hubbard Hamiltonian describes the dynamics of bosons in a double-well properly for small interaction energies compared to the level spacing of the trap potential \cite{Mil97}, such 
that only the two lowest-lying modes have to be considered. 
\\
According to the Bose-Hubbard Hamiltonian (\ref{eq_BH}) there are three qualitatively different regimes with respect to crucial features of energy spectrum and resulting dynamics \cite{Leg01,Pa01}. 
The regimes are easily explained by introducing the parameter $\Lambda=UN/T$ with the total number of particles in the system $N$. 
On the basis of $\Lambda$, the Rabi regime ($\Lambda < 1$) can be distinguished from the Josephson- ($1 < \Lambda \ll N^2$) and the Fock regime ($\Lambda \gg N^2$) \cite{Leg01}. 
\\
The Rabi regime can be identified with the non-interacting, classical limit $\Lambda \ll 1$, when the bosons move approximately independently of each other. 
The resulting spectrum is almost harmonic, such that after an initial tilt of the system the bosons oscillate with the well-known plasma frequency $\omega_p=2T\sqrt{1+\Lambda}$ \cite{Smer97,Gat07}. 
\\
The energy spectrum in the Fock regime only consists of doublets with a quasidegenerate symmetric and antisymmetric eigenstate. Thus, the dynamics of the mean population of the potential wells 
is extremely slow (self-trapping). 
\\
The Josephson regime combines the features of both spectra just discussed. Here, we distinguish the plasma-oscillating regime with $E<2NT$ and the self-trapping regime with $E>2NT$. 
In the former, the energy eigenstates correspond to an (anharmonic) oscillator spectrum and the population imbalance oscillates around zero, while in the latter regime the eigenstates 
appear as doublets leading to self-trapping again, and thus, to a mean population imbalance unequal zero. 
So, in the Josephson regime the dynamics depend crucially on the energy of the initial state, which is adjusted experimentally by initially tilting the potential \cite{Alb05}: The system 
is prepared in the ground state $\psi_0$ of the tilted potential, which is then quickly switched back to a symmetric potential and the bosons can evolve in time. 
\\
\\
Though the double-well system is a simple system, the resulting dynamics are interesting due to the interplay of tunnelling and interaction among the bosons. This
leads to sequences of oscillations, collapses, and revivals of the population imbalance and the relative phase, which we address later in this article.

\subsection{Classical description}
\label{sec_class_descr}

For a large total number of particles $N$ the Bose-Hubbard Hamiltonian (\ref{eq_BH}) can be simplified in the mean-field picture by replacing the annihilation and creation operators by complex numbers \cite{Smer97}
\begin{equation}
 \hat a_{i} \rightarrow \sqrt{n_i(t)}\exp(i \phi_i(t)) \ , 
\end{equation}
with the number of particles $n_i$ and phase $\phi_i$ in the $i$-th well. By introducing the 
population imbalance $j=(n_1-n_2)/2$ and the phase difference $\phi=\phi_1-\phi_2$ between the two wells, one finds the Hamilton function
\begin{equation}
 \label{eq_class_H}
 H(\phi,j)=\mathrm{const}+2Uj^2-2T\sqrt{(N/2)^2-j^2} \cos \phi \ ,
\end{equation}
which also arises from the mean-field Gross-Pitaevskii functional in the two-mode case \cite{Smer97}. 
The population imbalance and relative phase are canonically conjugate and the corresponding classical dynamics follow 
Hamilton's equations of motion
\begin{align}
  \frac{dj}{dt}&=-\frac{\partial H}{\partial \phi}=-2T \sqrt{(N/2)^2-j^2} 
\sin \phi
\nonumber
\\
\frac{d \phi}{dt}&=\frac{\partial H}{\partial j}=4Uj+\frac{2Tj \cos 
\phi}{\sqrt{(N/2)^2-j^2}} \ .
\label{eqofmotion}
\end{align}
The dynamics of the reduced quantities $j$ and $\phi$ have been studied by several groups in this ``classical picture'' \cite{Mil97,Rag99,Hol01}, 
in particular the differences between the classical and the quantum dynamics \cite{Ton05,Kra09,Jav10}. 
Though the purely classical description can clearly not describe the mentioned collapses and revivals of the population imbalance, it is able 
to give some indication of e.g. the transition from the plasma oscillating to the self-trapping regime and the order of magnitudes of the oscillation frequencies. 
Only recently the phase space region near the classical bifurcation was investigated experimentally with ultra cold $^{87}\mathrm{Rb}$ atoms \cite{Zib10}. 
On the other hand the classical description has been the basis of successful semiclassical investigations \cite{Fran00,Gra07,Sh08,Chu10,Nis10,Itin11,Paw11,Sim12} in the framework of WKB methods. 
Motivated by this preliminary work we want to move over to time-dependent semiclassical methods (namely the HK propagation) to describe the dynamics of the out-of-equilibrium ultracold bosons in the double-well potential. 
\subsection{TWA for the double-well system}
The Wigner representation can be written in the number-phase picture \cite{Hus10}. 
For a large number of particles, the differential equations can be truncated leading to a classical Liouville equation for the Wigner function
\begin{equation}
 \frac{\partial}{\partial t} W(\phi,j)=\left \{H(\phi,j),W(\phi,j)\right \} \ ,
\end{equation}
with the classical Hamilton function (\ref{eq_class_H}). Again, the difference to the purely classical treatment in section \ref{sec_class_descr} 
is that the distribution of the initial state is included: The whole Wigner function is propagated by means of the sampled single trajectories, 
that follow the classical equations (\ref{eqofmotion}) of section \ref{sec_class_descr}.

\subsection{Results} 
\label{sec_results}
\begin{figure}[htb]
\includegraphics[width=70mm]{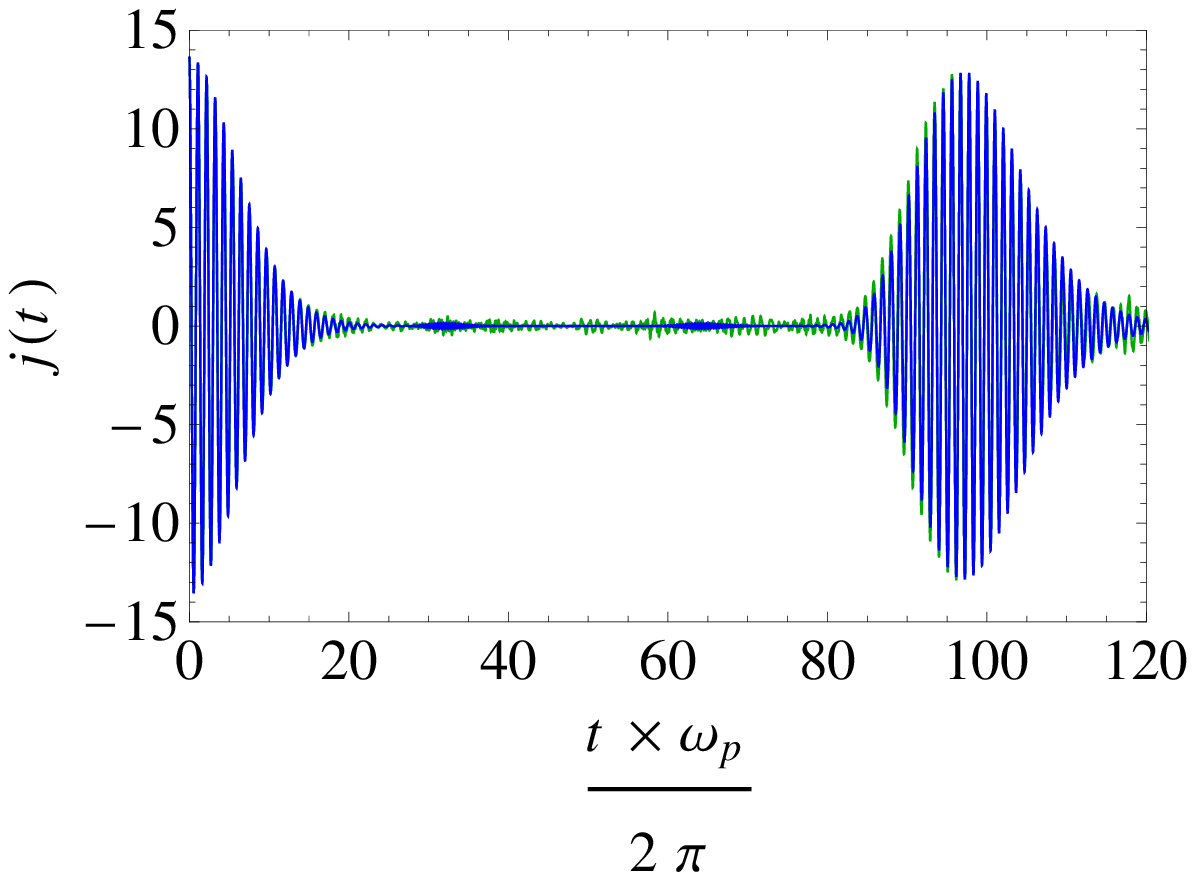}
\\
\includegraphics[width=70mm]{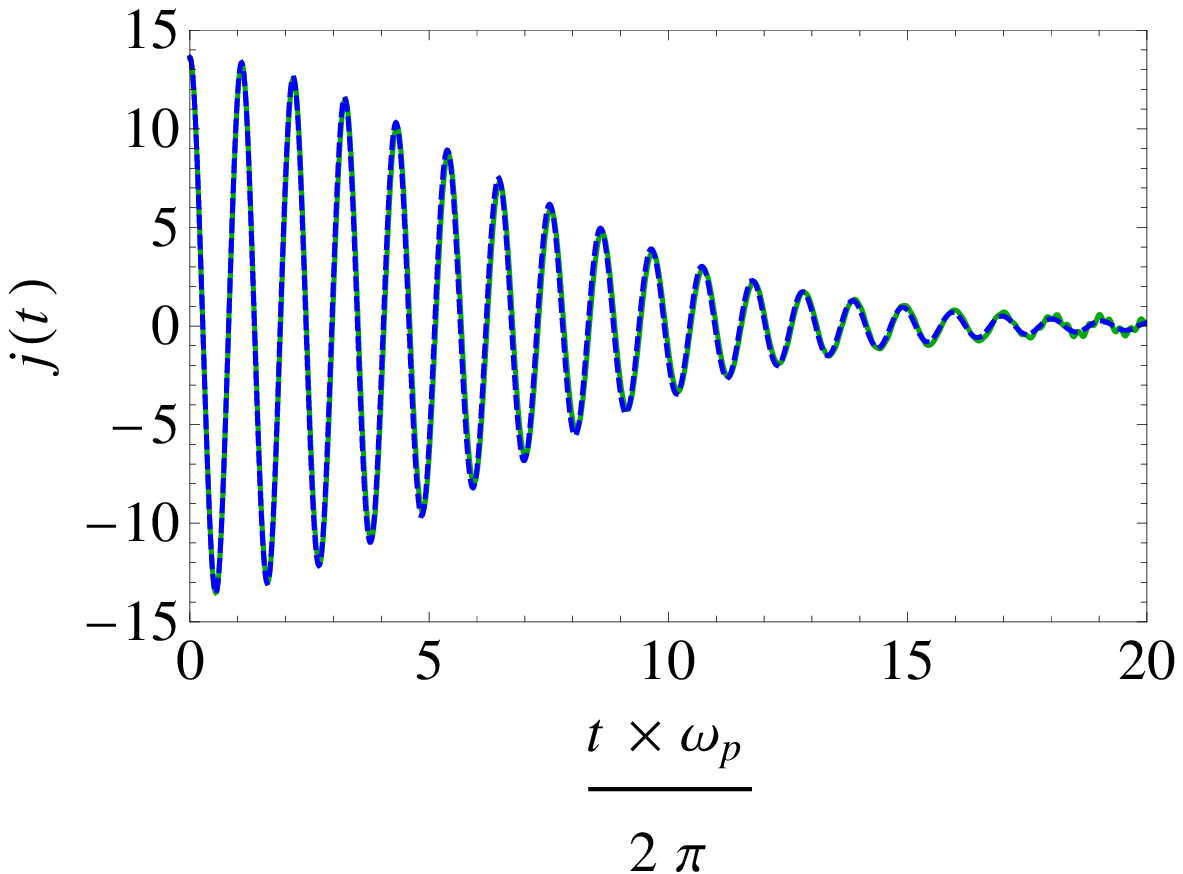}
\\
\includegraphics[width=70mm]{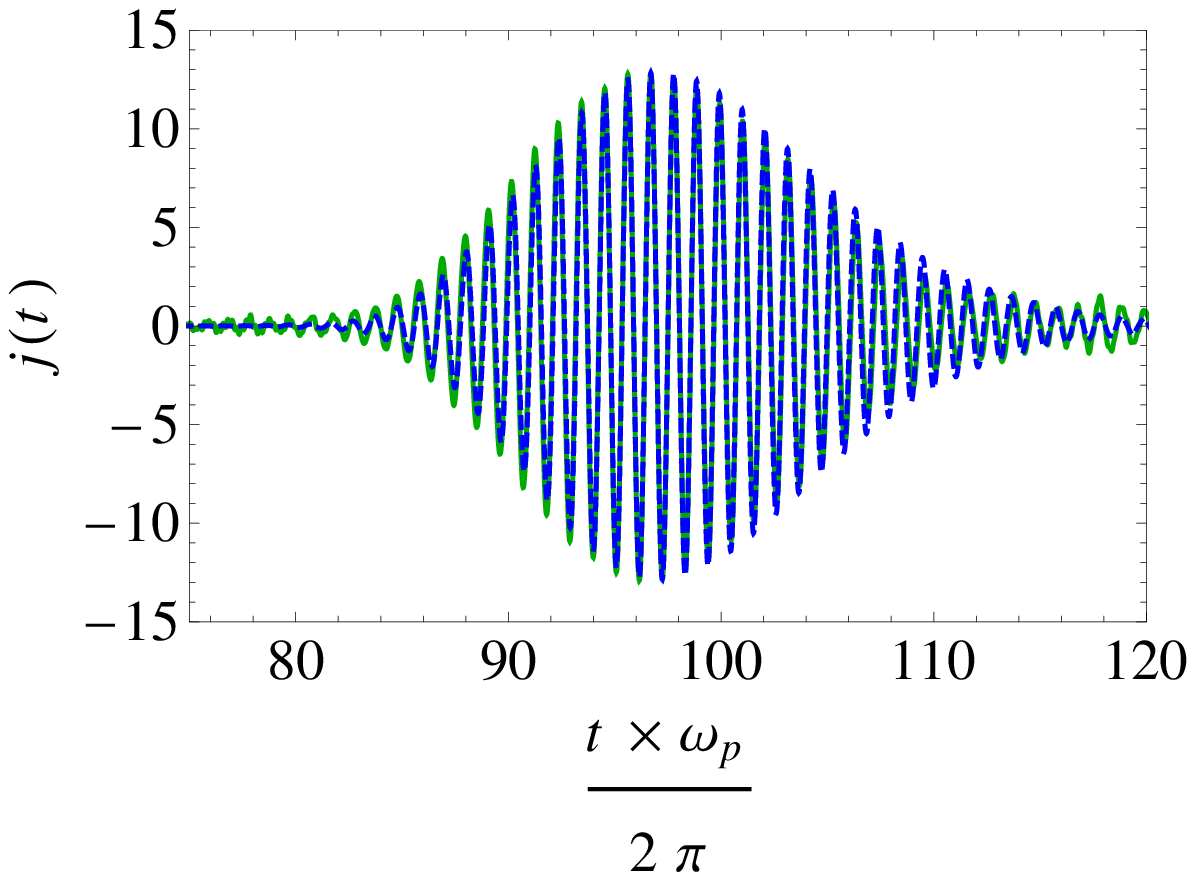}
\caption{Dynamics of the population imbalance in a double-well potential calculated numerically exactly (blue dashed line) and semiclassically by means of the HK propagator 
(green solid line) based on $10^4$ sampled trajectories for $\Lambda=10$, $N=100$, $T=10$ and initial displacement $j_0=14$ in different time frames in units of the plasma frequency $\omega_p$.  }
\label{fig_hk_jos}
\end{figure}

\begin{figure}[htb]
\includegraphics[width=70mm]{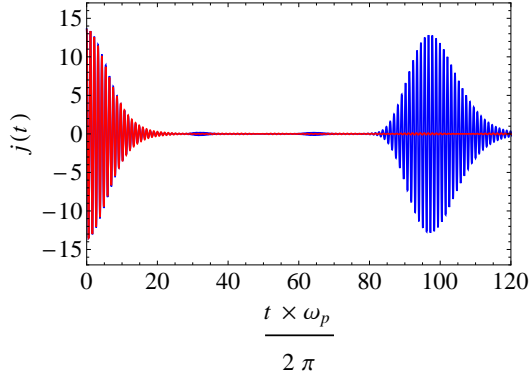}
\caption{Dynamics of the population imbalance in a double-well potential calculated numerically exactly (blue dashed line) and by means of the TWA 
(red solid line) based on $10^4$ sampled trajectories for $\Lambda=10$, $N=100$, $T=10$ and initial displacement $j_0=14$.  }
\label{fig_twa_jos}
\end{figure}

Since the initial state in the Heidelberg experiment \cite{Alb05} is almost a Gaussian in $(j,\phi)$ (being the ground state of the tilted system which can be calculated likewise semiclassically \cite{Sim12}), 
it seems to be natural to apply an initial value propagator to the system, namely the HK propagator. 
The canonically conjugate population imbalance and relative phase can be identified with position and momentum variables  
\begin{align}
 \phi \rightarrow q \ , \ \
j \rightarrow p \ .
\nonumber
\end{align}
Though $j$ and $\phi$ are finite variables, they can be considered as ordinary canonical variables in the limit $N\rightarrow \infty$. 
So, computing the HK propagator for the double-well system amounts to solving seven coupled equations of motion, namely Hamilton's equations of motion (\ref{eqofmotion}), 
the time evolution of the action, and four equations of motion for the entries of the monodromy matrix (\ref{eq_monodromy}), and solving the integral in (\ref{eq_hk}) by Monte-Carlo integration. 
Instead of sampling over the whole phase space, a Gaussian importance sampling is sufficient due to the Gaussian overlap (\ref{eq_overlap}) of the coherent states. 
So, the computational effort is increased only slightly compared to TWA methods, though the quantum coherences of the contributions of the different trajectories are included by means of the phases $\exp(iS/\hbar)$. 
\\
\\
Fig. \ref{fig_hk_jos} shows typical dynamics of the population imbalance in the plasma oscillating region of the Josephson regime, namely plasma oscillations around zero followed by a sequence 
of collapses and revivals. It can be seen that the semiclassical dynamics fit the quantum mechanical exact results very nicely and goes 
thus far beyond the TWA. The latter is displayed in  
Fig. \ref{fig_twa_jos}, showing the TWA dynamics for the same set of parameters. Obviously the TWA copes with the first collapse of the population imbalance, 
which is due to the whole bunch of considered trajectories 
representing the initial wave function in phase space. As a result of the nonlinearity brought in by the interaction of the atoms, the wave function smears out and gets 
distributed over the whole phase space after some time leading to the collapse of the population imbalance (this can be seen in a quantum phase space picture e.g. in Ref. \cite{Mah05}). 
Since the TWA neglects quantum coherences, 
the quantum revivals caused by interference effects cannot be reproduced. 

\begin{figure}[htb]
\includegraphics[width=70mm]{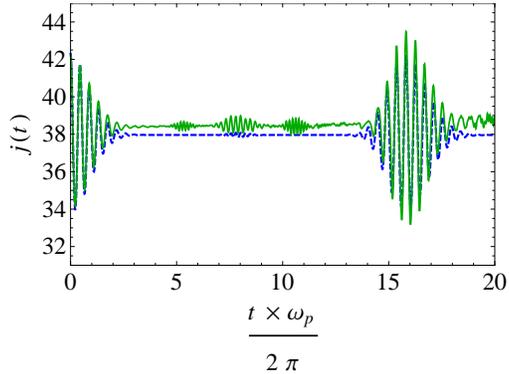}
\caption{Self-trapping dynamics of the population imbalance in a double-well potential numerically exact (blue dashed line) and calculated semiclassically by means of the HK propagator 
(green solid line) based on $10^4$ sampled trajectories for $\Lambda=100$, $N=100$, $T=10$ and initial displacement $j_0\approx 40$.  }
\label{fig_hk_st}
\end{figure}

It comes as no surprise that the dynamics of the population imbalance in the plasma oscillating regime can be described well by means of the semiclassical HK propagator. 
Even an analytical expression for the semiclassical dynamics was found for this regime \cite{Sim12}. However, it turns out that the dynamics in the 
self-trapping regime can be reproduced too, as shown in Fig. \ref{fig_hk_st}. Only the little intermediate revivals \cite{Sim12} are overestimated by the HK method. 
Also, the value of the population imbalance in the temporary stationary state is slightly shifted to the exact one, which can be traced back to the self-trapping properties of the 
phase space. In the latter regime the phase is ever growing leading to an unstable norm of the HK wave function. Despite renormalization, the semiclassical result is slightly shifted compared to 
the exact one. Those difficulties aside, the semiclassical methods reproduce all the dynamical characteristics of the population imbalance exactly. 

\begin{figure}[htb]
\includegraphics[width=60mm]{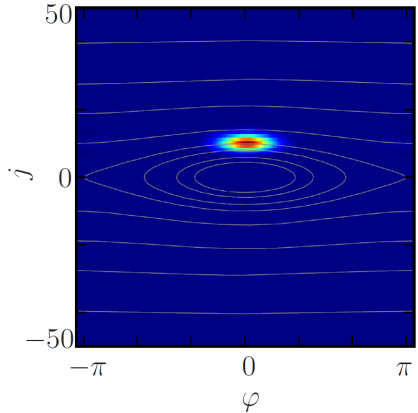}
\caption{Wigner function of an initial wave function in phase space for $\Lambda=100$, $N=100$, $T=10$ and $j_0\approx 10$. The lines mark the 
equipotential curves. }
\label{fig_wig_phase_space}
\end{figure}

\begin{figure}[htb]
\includegraphics[width=70mm]{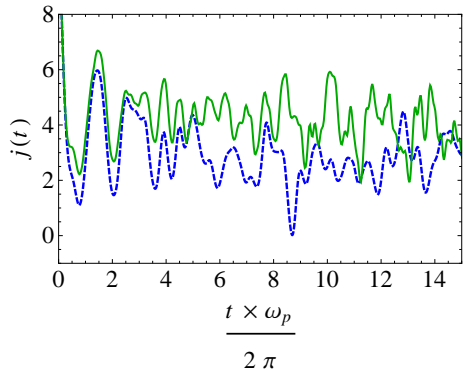}
\caption{Dynamics of the population imbalance in a double-well potential numerically exact (blue dashed line) and calculated by means of the HK propagator 
(green solid line) based on $10^4$ sampled trajectories for $\Lambda=100$, $N=100$, $T=10$ and initial displacement $j_0\approx 10$  in units of the plasma frequency $\omega_p$.  }
\label{fig_hk_sep}
\end{figure}

Next we want to investigate the case where the initial wave function is located in the plasma oscillating and the self-trapping regime at the same time, 
it is thus spread over the separatrix in phase space, which separates plasma-oscillating and self-trapping regime. 
Such an initial condition (shown in phase space in Fig. \ref{fig_wig_phase_space}) leads to irregular dynamics, as shown in Fig. \ref{fig_hk_sep}. 
It turns out that the HK propagator is able to reproduce the respective dynamics only on very short timescales. 
This can be traced back to the classical phase space again: It has been shown \cite{Shc07} that the quantum mechanical time evolution discriminates between 
stable and unstable classical fixed points. It follows the classical dynamics close to stable fixed points and diverges around unstable ones. 
Since the classical dynamics form the basis of the semiclassical approaches, it is clear that the exact quantum dynamics cannot be described close to unstable hyperbolic fixed points. 

\begin{figure}[htb]
\includegraphics[width=70mm]{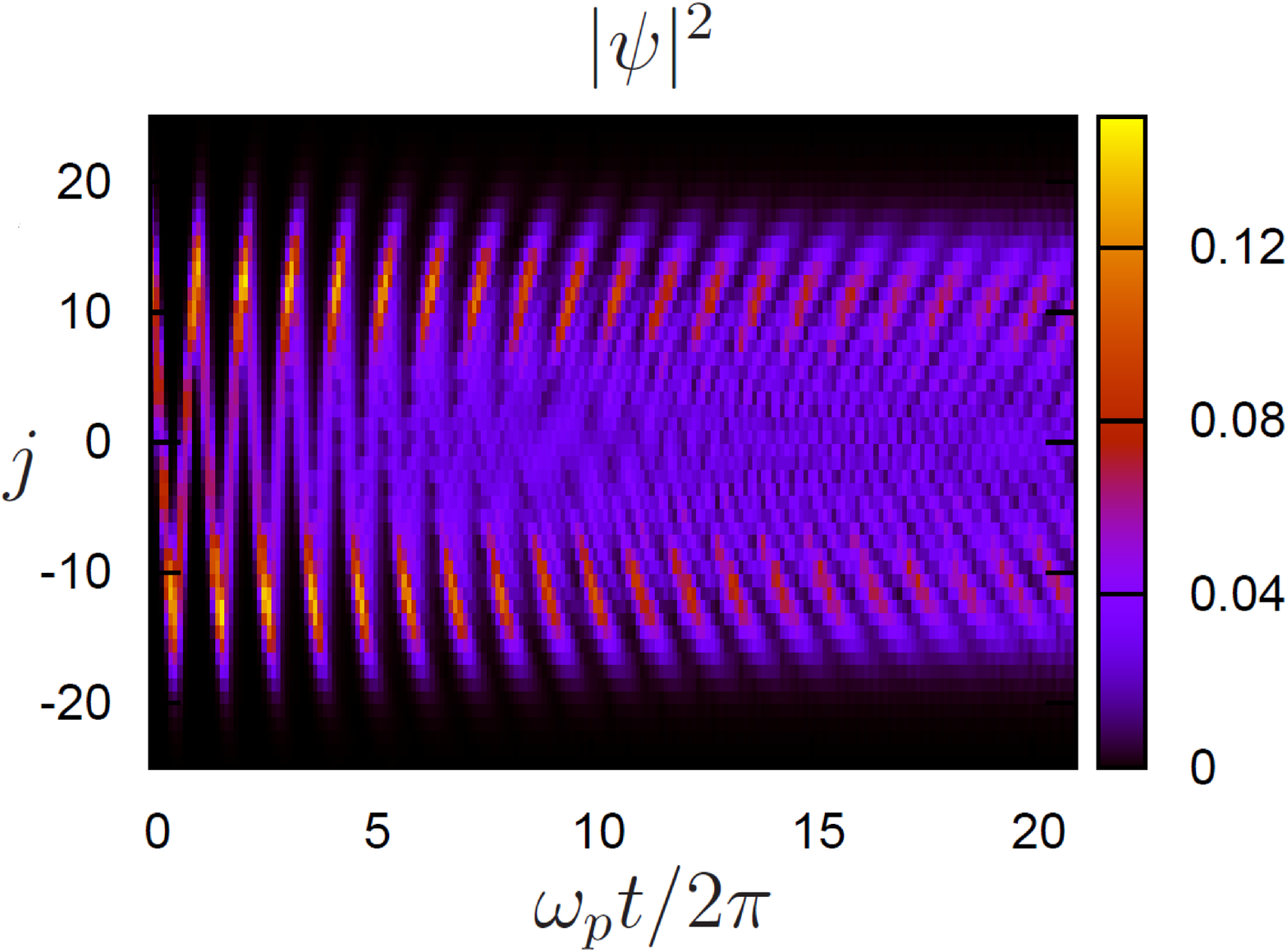}
\includegraphics[width=70mm]{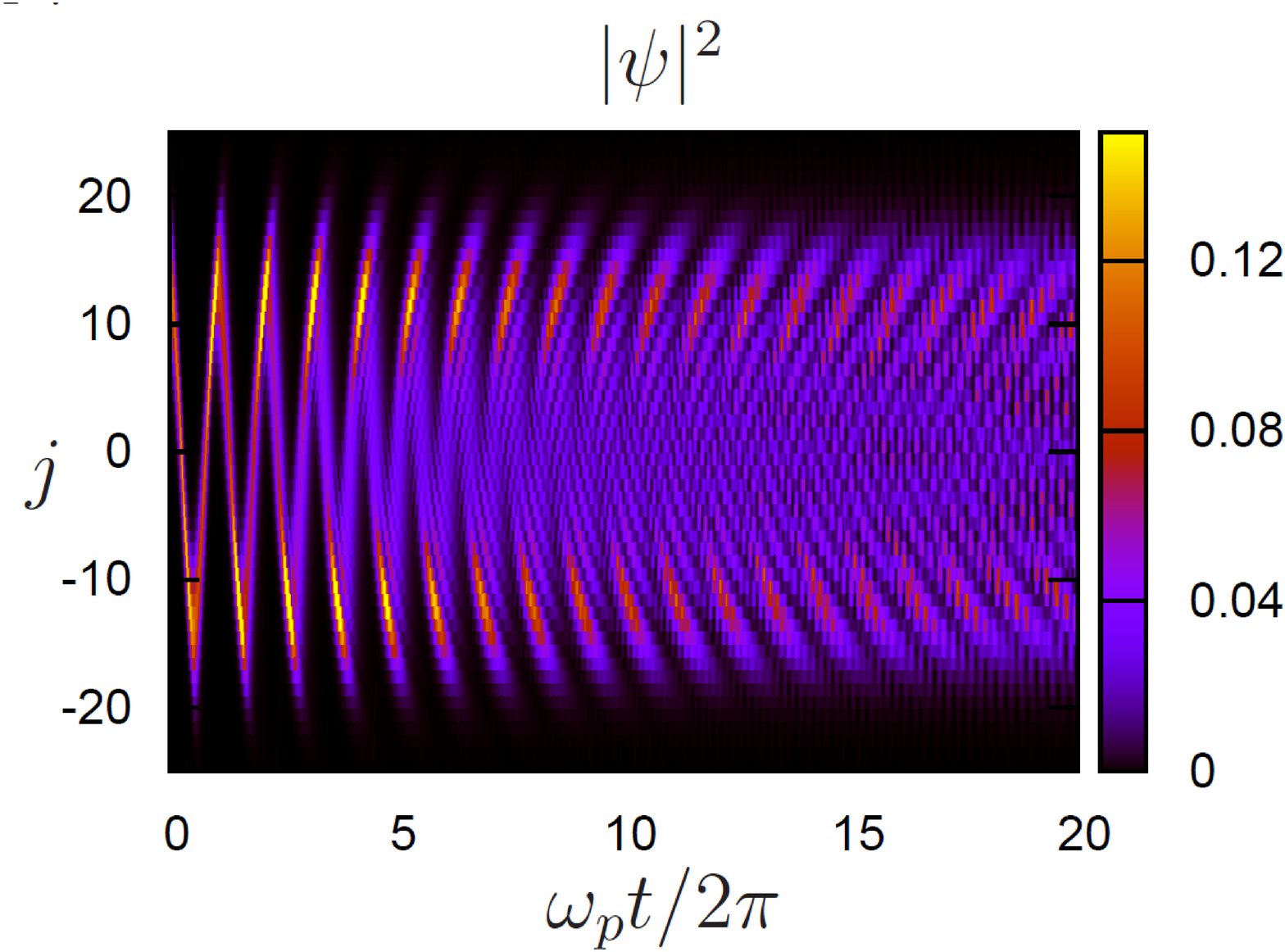}
\caption{Dynamics of the wave function in momentum space in a double-well potential numerically exact (left) and calculated by means of the HK propagator 
(right) based on $10^4$ sampled trajectories for $\Lambda=10$, $N=100$, $T=10$ and initial displacement $j_0=14$ in units of the plasma frequency $\omega_p$.  }
\label{fig_wf}
\end{figure}

Finally, we want to mention that the HK propagation does not only allow for an investigation of reduced quantities such as the population imbalance as most other semiclassical descriptions. 
Instead, it is possible to compute the whole wave function as shown in Fig. \ref{fig_wf} for the same parameters as in  Fig. \ref{fig_hk_jos}. 
For the sake of clarity, the timescale is chosen to be roughly the collapse time. By means of the colour code the decay of the population imbalance is observable. 
Obviously, the semiclassical wave function agrees surprisingly accurately with the exact one. 

To summarize, the semiclassical HK propagation is able to reproduce the dynamics of the atomic gas in the double-well potential in the framework of the 
two-mode approximation going clearly beyond the TWA, even though the computational effort is similar compared to the TWA -- only the phase and stability of the respective trajectory has to be calculated additionally. 

\section{The triple-well system}
\begin{figure}[htb]
\includegraphics[width=70mm]{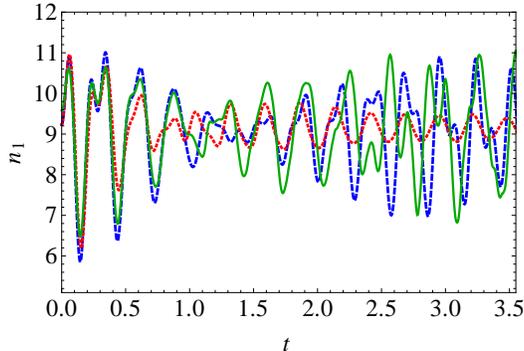}
\caption{Dynamics of the number of particles in the first well $n_1$ for a triple-well potential calculated numerically exactly (blue dashed), with truncated Wigner approximation (red dotted) and by means of the HK propagator 
(green solid) based on $2 \times 10^4$ sampled trajectories for $N=30$, $T=10$, $U=1$ and $\delta=10$.  }
\label{fig_triple}
\end{figure}

Since our goal is to describe the dynamics of ultracold bosons in larger optical lattices we 
proceed to three potential wells. The triple-well system is not integrable anymore leading to an even 
richer dynamical behaviour than the double-well setup \cite{Nem00,Fra01,Fra03}. 
Despite the simplicity of the system, the resulting dynamics is affected by an inner instability accompanied by chaotic behaviour \cite{Fra03,Tho03,Mos06}, 
originating from the nonlinear dynamics in a four dimensional phase space leading to non-integrability. Depending on the 
system parameters there are different regimes: Integrable subsystems lead to regular dynamics and self-trapping, while there occurs chaotic dynamics 
out of the regular islands. The system has been investigated comprehensively by several groups \cite{Nem00,Fra01,Fra03,Tho03,Mos06}. 
Roughly speaking again, the dynamics in the triple-well system depends sensitively on the parameter $\Lambda$ and the initial imbalance 
between the wells. 
\\
\\
Similarly to the double-well system, the basis for the semiclassical and the truncated Wigner approximation is the 
Bose-Hubbard Hamiltonian, but extended by a third mode
\begin{equation}
 \hat H_{BH,3}=U(\hat a_1^\dagger \hat a_1^\dagger \hat a_1 \hat a_1 + \hat a_2^\dagger \hat a_2 ^\dagger \hat a_2 \hat a_2 + \hat a_3^\dagger \hat a_3^\dagger \hat a_3 \hat a_3 )
-T(\hat a_1^\dagger \hat a_2 + \hat a _2^\dagger \hat a_1 + \hat a_2^\dagger \hat a_3 + \hat a_3^\dagger \hat a_2) \ . 
\end{equation}
Once more we choose the ground state of the tilted system with Hamiltonian $\hat H = \hat H_{BH,3} + \delta(\hat n_1 - \hat n_2)$ as initial state. 
The classical dynamics can be found in accordance to section \ref{sec_class_descr} 
leading to four coupled equations of motion for the particle numbers in two of the wells, say $n_1$ and $n_2$, 
and the respective phase differences $\varphi_1=\phi_1-\phi_2$ and $\varphi_2=\phi_2-\phi_3$ \cite{Mos06}. Even though the classical approximation is quite rough, it can 
contribute to a qualitative understanding of the dynamics -- the type of the classical phase space determines the nature of the quantum dynamics \cite{Fra03,Mos06}. 
\\
\\
It turns out that the initial regular dynamics (for small initial imbalance and small $\Lambda$) are quite well approximated by the TWA (thus there is no need for the HK propagation for short times). 
On the other hand, for increasing $\Lambda$ the phase space becomes more and more chaotic reducing the applicability of the HK propagation. As we stated before, the 
HK prefactor (\ref{eq_HK_pref}) is a measure for the stability of the trajectories. The contributions of 
chaotic trajectories are growing exponentially in time \cite{Kay942}, namely with the respective Lyapunov exponent \cite{Mey86}. 
In practice this implies that only a few highly unstable chaotic trajectories can dominate the semiclassical wave function completely \cite{Kay94}, 
implicating an exponentially growing corresponding norm in time. 
This leads to problems with convergence and errors in the calculations. 
Usually, those kinds of trajectories are sorted out if their occurrence is in the single-digit percentage range \cite{Kay942}, 
but obviously this procedure implicates an inaccuracy of the semiclassical results. 
Fig. \ref{fig_triple} shows such a dynamics. It can be seen, that neither the HK propagation nor the TWA is able to describe the exact dynamics on a long time scale precisely. 
However, the semiclassical results clearly follow the proper dynamics for a longer time than TWA. 
In contrast, the TWA results decay fast and as discussed in the previous section \ref{sec_results}, the TWA cannot describe any revivals due to the lack of phase information. 
However, the semiclassical treatment generates the right envelope of the revivals.

\section{Conclusion}
We study the dynamics of ultracold bosons in a double- and triple-well potential by means of the semiclassical initial value HK propagator in the framework of the 
two- and three-mode approximation and compare the results to the frequently applied truncated Wigner approximation. For both descriptions, only ``classical information'' is required. 
Concerning the double-well system we find that the semiclassical propagation is able to reproduce the exact results for most parameter regimes, except for 
initial wave functions lying on the separatrix in phase space. Hence, semiclassics clearly goes beyond the TWA. The latter is only able to describe the dynamics 
on the timescale of the first collapse of the population imbalance correctly, since quantum coherences are neglected. Depending 
on the parameter $\Lambda$ this timescale can be of the order of only a few oscillations, as can be seen in Fig. \ref{fig_hk_st}, 
making the HK propagation an essential improvement compared to the TWA. 
For the irregular triple-well system, where the TWA seizes to be a valid description even for short times, we show that there are cases where the HK propagation also describes the correct 
dynamics on longer timescales. However, for longer times the problems with unstable trajectories also limit the HK propagation. 
Switching to a position-momentum representation instead of the number-phase representation might improve the results. 
 
\section*{Acknowledgments}
We thank the Herman Kluk experts Frank Großmann and Werner Koch for fruitful discussions. 
L. S. acknowledges support from the International Max Planck Research
School (IMPRS), Dresden.


\begin{thebibliography}{99}

\bibitem{Polk11}
A.~ Polkovnikov, K.~ Sengupta, A.~ Silva, and M.~ Vengalattore, 
Rev. Mod. Phys. \textbf{83}, 863 (2011). 

\bibitem{Kenn12}
M.~P.~Kennett, ISRN Condensed Matter Physics \textbf{2013}, 393616 (2013). 

\bibitem{Trotz12}
S.~ Trotzky, Y-A.~ Chen, A.~ Flesch, I.~P.~ McCulloch, U.~ Schollwöck, J.~ Eisert, and I.~ Bloch, 
Nature Physics \textbf{8}, 325 (2012). 

\bibitem{Jak98}
D.~ Jaksch, C.~ Bruder, J.~I.~ Cirac, C.~W.~ Gardiner, and P.~ Zoller, 
Phys. Rev. Lett. \textbf{81}, 3108 (1998). 

\bibitem{String03}
L.~P.~ Pitaevskii and S.~ Stringari, 
\textit{Bose-Einstein Condensation}, 
(International Series of Monographs on Physics, Clarendon Press, 2003). 


\bibitem{Berry72}
M.~V.~ Berry and K.~E.~ Mount, 
Reps. Prog. Phys. \textbf{35}, 315 (1972). 


\bibitem{VV28}
J.~H.~ van Vleck, 
Proc. Acad. Nat. Sci. USA \textbf{14}, 178 (1928). 

\bibitem{Gutz90}
M.~C.~ Gutzwiller, \textit{Chaos in Classical and Quantum Mechanics}, 
(Springer Series on Interdisciplinary Applied Mathematics, Springer, 1990). 


\bibitem{Frank}
F.~ Großmann, \textit{Theoretical Femtosecond Physics: Atoms and Molecules in Strong Laser Fields}, 
(Springer Series on Atomic, Optical, and Plasma Physics, Springer, 2008). 

\bibitem{Var01}
A.~ Vardi, and J.~R.~ Anglin,
Phys. Rev. Lett. \textbf{86}, 568 (2001).

\bibitem{Ang01}
J.~R.~ Anglin, and A.~ Vardi,
Phys. Rev. A \textbf{64}, 013605 (2001).

\bibitem{Fran00}
R.~ Franzosi, V.~ Penna and R.~ Zecchina,
Int. Jour. of Mod. Phys. B \textbf{14}, 943 (2000).

\bibitem{Gra07}
E.~M.~ Graefe, and H.~J.~ Korsch,
Phys. Rev. A \textbf{76}, 032116 (2007).

\bibitem{Sh08}
V.~S.~ Shchesnovich, and M.~ Trippenbach,
Phys. Rev. A \textbf{78}, 023611 (2008).

\bibitem{Chu10}
M.~ Chuchem, K.~ Smith-Mannschott, M.~ Hiller, T.~ Kottos, A.~ Vardi, 
and D.~ Cohen,
Phys. Rev. A \textbf{82}, 053617 (2010).

\bibitem{Nis10}
F.~ Nissen, and J.~ Keeling,
Phys. Rev. A \textbf{81}, 063628 (2010).

\bibitem{Itin11}
A.~P.~ Itin and P.~ Schmelcher, 
Phys. Rev. A \textbf{84}, 063609 (2011). 

\bibitem{Paw11}
K.~ Pawlowski, P.~ Zin, K.~ Rzazewski, and M.~ Trippenbach,
Phys. Rev. A \textbf{83}, 033606 (2011).

\bibitem{Sim12}
L.~ Simon and W.~T.~ Strunz, 
Phys. Rev. A \textbf{86}, 053625 (2012). 

\bibitem{Lieb00}
E.~H.~ Lieb, R.~ Seiringer, and J.~ Yngvasion, 
Phys. Rev. A \textbf{61}, 043602 (2000). 

\bibitem{HK84}
M.~F.~ Herman and E.~ Kluk, 
Chem. Phys. \textbf{91}, 27 (1984). 

\bibitem{Hel81}
E.~J.~ Heller, J. Chem. Phys. \textbf{75}, 2923 (1981). 

\bibitem{Kay94}
K.~G.~ Kay, J. Chem. Phys. \textbf{100}, 4377 (1994). 

\bibitem{Kay942}
K.~G.~ Kay, J. Chem. Phys. \textbf{101}, 2250 (1994). 

\bibitem{Juan13}
T.~ Engl, J.~ Dujardin, A.~ Argülles, P.~ Schlagheck, K.~ Richter, and J.~D.~ Urbina, 
arXiv:1306.3169v1 (2013). 

\bibitem{Steel98}
M.~J.~ Steel, M.~K.~ Olsen, L.~I.~ Plimak, P.~D.~ Drummond, S.~M.~ Tan, M.~J.~ Collett, D.~F.~ Walls, and R.~ Graham, 
Phys. Rev. A \textbf{58}, 4824 (1998). 

\bibitem{Sin02}
A.~ Sinatra, C.~ Lobo, and Y.~ Castin, 
J. Phys. B: At. Mol Opt. Phys. \textbf{35}, 3599 (2002). 

\bibitem{Blak09}
P.~B.~ Blakie, A.~S.~ Bradley, M.~J.~ Davis, R.~J.~ Ballagh, and C.~W.~ Gardiner, 
Adv. Phys. \textbf{57}, 363 (2008). 

\bibitem{Gat07}
R.~ Gati, and M.~K.~ Oberthaler,
At. Mol. Opt. Phys. \textbf{40}, R61 (2007).

\bibitem{Jak05}
D.~ Jaksch and P.~ Zoller, 
Ann. Phys. \textbf{315}, 52 (2005). 

\bibitem{Mil97}
G.~J.~ Milburn, J.~ Corney, E.~M.~ Wright, and
D.~F.~ Walls,
Phys. Rev. A \textbf{55}, 4318 (1997).

\bibitem{Leg01}
A.~ Leggett, Rev. Mod. Phys. \textbf{73}, 307 (2001).

\bibitem{Pa01}
G.~S.~ Paraoanu, S.~ Kohler, F.~ Sols, and A.~ Leggett,
At. Mol. Opt. Phys. \textbf{34}, 4689 (2001).


\bibitem{Smer97}
A.~ Smerzi, S.~ Fantoni, S.~ Giovanazzi, and S.~R.~Shenoy,
Phys. Rev. Lett. \textbf{79}, 4950 (1997).

\bibitem{Alb05}
M.~ Albiez, R.~ Gati, J.~ F\"olling, S.~ Hunsmann, M.~ Cristiani, and 
M.~K.~ Oberthaler,
Phys. Rev. Lett \textbf{95}, 010402 (2005).

\bibitem{Rag99}
S.~ Raghavan, A.~ Smerzi, S.~ Fantoni, and S.~R.~ Shenoy,
Phys. Rev. A \textbf{59}, 620 (1999).

\bibitem{Hol01}
M.~ Holthaus, and S.~ Stenholm,
Eur. Phys. J. B \textbf{20}, 451 (2001).

\bibitem{Ton05}
A.~P.~ Tonel, J.~ Links, and A.~ Foerster,
J. Phys. A: Math. Gen. \textbf{38}, 6879 (2005).

\bibitem{Kra09}
G.~J.~ Krahn, and D.~H.~J.~ O'Dell,
J. Phys. B  \textbf{42}, 205501 (2009).

\bibitem{Jav10}
J.~ Javanainen,
Phys. Rev. A \textbf{81}, 051602(R) (2010).


\bibitem{Zib10}
T.~ Zibold, E.~ Nicklas, C.~ Gross and M.~K.~ Oberthaler,
Phys. Rev. Lett. \textbf{105}, 204101 (2010).

\bibitem{Hus10}
M.~R.~ Hush, A.~R.~R. Carvalho, and J.~J.~ Hope, 
Phys. Rev. A \textbf{81}, 033852 (2010). 

\bibitem{Mah05}
K.~W.~ Mahmud, H.~ Perry, and W.~P.~ Reinhardt,
Phys. Rev. A \textbf{71}, 023615 (2005).

\bibitem{Shc07}
V.~S.~ Shchesnovich and V.~V.~ Konotop, 
Phys. Rev. A \textbf{75}, 063628 (2007). 


\bibitem{Fra03}
R.~ Franzosi and V.~ Penna, 
Phys. Rev. E \textbf{67}, 046227 (2003). 

\bibitem{Nem00} 
K.~ Nemoto, C.~A.~ Holmes, G.~J.~ Milburn, and W.~J.~ Munro, 
Phys. Rev. A \textbf{63}, 013604 (2000). 

\bibitem{Fra01}
R.~ Franzosi and V.~ Penna, 
Phys. Rev. A \textbf{65}, 013601 (2001). 

\bibitem{Tho03}
Q.~ Thommen, J.~C.~ Garreau, and V.~ Zehnle, 
Phys. Rev. Lett. \textbf{91}, 210405 (2003). 

\bibitem{Mos06}
S.~ Mossmann and C.~ Jung, 
Phys. Rev. A \textbf{74}, 033601 (2006). 

\bibitem{Mey86}
H.~D.~ Meyer, 
J. Chem. Phys. \textbf{84}, 3147 (1986). 


%
%
%
%
%
%

%
%
%
%
%
%
%
%
%
%
%
%
%
%
%
%
%
%
%
%
%
%
%
%
%
%
%
%


\end{thebibliography}
\end{document}